\documentclass[12pt]{iopart}
\input epsf

\begin{document}

\title[Viscous slowing-down of supercooled liquids]{The  viscous
slowing   down   of     supercooled   liquids as       a
temperature-controlled superArrhenius  activated process: a description
in terms of frustration-limited domains}

\author{
Gilles Tarjus$^*$, Daniel Kivelson$^{**}$, and Pascal Viot$^*$}

\address{$^*$ Laboratoire de Physique Th{\'e}orique des Liquides,
Universit{\'e} Pierre  et Marie Curie,  4 Place Jussieu, 75252 Paris cedex
05, France\\ $^{**}$ Department of Chemistry and Biochemistry, University
of California, Los Angeles, CA 90095, USA}


\begin{abstract}

We propose that  the salient feature to  be explained  about the glass
transition  of  supercooled    liquids is the   temperature-controlled
superArrhenius  activated  nature of the    viscous slowing down, more
strikingly  seen in weakly-bonded, fragile   systems.  In the light of
this observation,  the  relevance  of  simple  models   of spherically
interacting particles  and   that  of  models  based   on  free-volume
congested dynamics are  questioned. Finally, we  discuss how the  main
aspects of   the phenomenology of  supercooled  liquids, including the
crossover from Arrhenius to superArrhenius  activated behavior and the
heterogeneous character of the $\alpha$ relaxation, can be described by an
approach based on frustration-limited domains.
\end{abstract}
\submitted{{\noindent \it }}

\section{Introduction}
What is there to be explained  about supercooled liquids and the glass
transition?   The lack of agreement  over the  answer to this question
certainly contributes   to  the multiplicity of  proposed   models and
theories,  many  of which   seem  orthogonal to  each other.   Our own
starting point,  further motivated and  discussed in this  article, is
that  the   most  distinctive  feature  of  supercooled    liquids, in
particular of  the so-called "fragile" liquids on  which we  focus, is
the  stupendous  continuous  increase     of viscosity  ($\eta$)     and
$\alpha$-relaxation time  ($\tau_\alpha$) with decreasing  temperature (some  15
order-of-magnitude increase over the range over which the substance is
liquid,  i.  e., a factor-of-two change  in  temperature), an increase
that is characterized for most liquids (with the possible exception of
strong network-forming systems) by a crossover  at a temperature $T^*$
from a roughly  Arrhenius temperature dependence  of $\eta$ and $\tau_\alpha $
for $T>T^*$ to  a faster-than- Arrhenius, or superArrhenius dependence
for $T<T^*$.  The viscous slowing down appears to be best described as
a thermally activated process, whose evolution is primarily controlled
by temperature  and   not by  volume    or density (at least  at   low
pressure). It is of particular interest  that the effective activation
free energy for viscous flow and $\alpha$ relaxation in {\it weakly-bonded
systems}, such as  the  much  studied ortho-terphenyl,  when  measured
relative to the energy characteristic  of thermal fluctuations $k_BT$,
varies by a  factor of $5$ or $6$  between the melting point $T_m$ and
the  glass ``transition'' temperature  $T_g$  at which it reaches  the
high value of about   $40$, a value  that is   comparable to  that  of
strongly-bonded systems such as  silica glasses.  Concomitant to these
variations  and  adding to  the puzzle   is  the absence of associated
singular     behavior or  strong     signature   in    the  structural
data\cite{EAN96}.    Whereas  mounting  evidence   has  been  recently
provided that shows the slow dynamics in deeply supercooled liquids to
be heterogeneous,  with  spatial heterogeneities  (containing of   the
order of $10^2$ molecules around $T_g$)  being largely responsible for
the    nonexponential character of the   $\alpha$   relaxation and for the
decoupling of translational diffusion  time scales from  viscosity and
rotational  relaxation  times,\cite{EAN96,S99} direct diffraction data
do  not  seem to  reflect the  existence of  supermolecular structural
correlations.   These   are the  phenomena   that we find particularly
intriguing about  the  glass transition,  and  our  viewpoint  is that
models, theories,   and  computer  simulations should   be  ultimately
evaluated in terms of their ability to explain and describe these main
features.

In the   following, we  first    present  some experimental   evidence
supporting  the  description    of  the   viscous   slowing  down   of
glass-forming liquids as  a temperature- controlled activated  process
crossing over to superArrhenius    $T$-dependence in the   supercooled
regime. The emphasis is placed  on weakly-bonded, ``fragile'' liquids.
In the light of these observations,  the relevance of simple atomistic
models used in most computer simulations  is addressed next.  Finally,
we summarize our theoretical approach of  supercooled liquids based on
frustration-limited  domains and    discuss its  main ingredients  and
predictions.

\section{The viscous slowing down as a $T$-controlled superArrhenius activated process}

The dramatic increase   of  viscosity and  $\alpha$-relaxation   time with
decreasing temperature in supercooled liquids  calls for a logarithmic
representation that is traditionally displayed in an Arrhenius type of
diagram, with $\log_{10}(\eta$ or $\tau_\alpha )$ as a function of $1/T$,
or as advocated by Angell\cite{ANW91} as  a function of $T_g/T$. As is
already well known  and is  illustrated in  Fig.1a,  systems forming a
network     of strong intermolecular  bonds     such  as $GeO_2$   are
characterized by a  rather linear variation  in $1/T$,  i.   e., by an
Arrhenius temperature  dependence, while   the  other liquids  show  a
marked upward curvature  which represents a faster-than-Arrhenius,  or
superArrhenius,   temperature  dependence.   This  latter feature   is
particularly noticeable for  weakly-bonded,  ``fragile'' liquids, such
as ortho-terphenyl\cite{ANW91}. A different way of presenting the phenomenon is
to plot  the effective  activation  free energy for a   relaxation and
viscosity, E(T), specified by
\begin{equation}
\tau_\alpha =\tau_{\alpha,\infty}\exp\left(\frac{E(T)}{k_BT}\right),
\end{equation}
where $k_B$ is the  Boltzmann constant and  $ \tau_{\alpha,\infty}$ is a  high-$T$
relaxation  time  taken  as independent    of temperature  (a  similar
equation holds for  the viscosity). This  is illustrated in  Fig.  1b,
where the temperature is scaled for each liquid  to a temperature $T^*$
above which the dependence is  roughly Arrhenius-like, i.  e.,  $E(T)$
is,  to a good approximation,  constant.  The determination of $T^*$ is
subject  to  some  uncertainty,\cite{KTZK96,SFR95}  but  the procedure
stresses the crossover from  Arrhenius-like to superArrhenius behavior
that is typical of most glassforming  liquids.  A first observation is
that $E(T)/k_BT$ is large compared to $1$ (it is about $40$ at $T_g$),
which strongly suggests that the $\alpha$  relaxation and the viscous flow
in supercooled  liquids  are  best described   as  thermally activated
processes. The paradigm for an activated process is that of a chemical
reaction where the activation barrier, typically more than $100 k_BT$,
is reasonably  independent   of temperature (indicating  an  Arrhenius
behavior) and a reaction path involving a small set of coordinates may
be well identified.  Relaxation in  strong network-forming systems, in
which   molecular transport requires  that  enough  thermal energy  be
provided to break the strong  intermolecular bonds, might be  expected
to  fit this simple  picture.  However, the activated relaxation found
in weakly-bonded, fragile  liquids   is astonishing both because   the
effective activation free energy increases markedly as the temperature
is lowered towards $T_g$ (a factor of $3$, or in units of $k_BT$ a factor
of $6$, in the case of  ortho-terphenyl) and because it reaches values
that are  much bigger  than  the typical  energy associated  with  the
intermolecular bonding. What  are then the  source and the explanation
of this superArrhenius activated behavior? They cannot be described by
conventional transition-state theories used in the context of chemical
reactions nor  by standard theories of  critical  slowing down  in the
vicinity of a usual second-order phase transition.  In connection with
this  last statement we note  that  the slowing  down of dynamics when
approaching   a    critical point    (here,  a  hypothetical  singular
temperature    below  $T_g$  as     predicted  for  instance  by   the
Vogel-Fulcher-Tammann formula) is usually characterized by a power-law
growth of the relaxation time,  which does not appear  to fit the data
on supercooled   liquids    with  reasonable  low    values of     the
exponent.  Unusually strong slowing   down, with exponentially growing
times, is however  found near  the critical  point of some  disordered
systems like the  random field Ising  model and is known as  ``activated
dynamic scaling\cite{FGK88}.  

An additional   characteristic  of the  slowing   down  of   the  $\alpha$
relaxation when approaching the glass transition is that it appears to
be predominantly controlled by  temperature and not  by density.  What
is   meant by this   statement,  and what  are   the consequences  for
potential   theoretical explanations?    Experiments on    supercooled
liquids are generally carried out at constant pressure, overwhelmingly
at atmospheric pressure.  As a result, a  decrease of temperature also
produces an  increase in density (typically a  $5\%$ increase over the
liquid range  at   $P=1atm$).  The question  is  whether  the dramatic
slowing  down and  the  above  mentioned crossover to   superArrhenius
behavior observed   at atmospheric pressure are  primarily  due  to an
intrinsic temperature effect or to  the influence of density, hence of
packing.  This  can  be answered by   using  pressure studies of  $\alpha$
relaxation and viscosity in  glassforming liquids.  A pictorial way of
addressing the  problem is  to plot $\log_{10}(\eta$  or $\tau_\alpha  )$ as a
function of $1/T$  at constant density $\rho$  and as a function of $\rho$
at constant $ T$, for  the range of  densities that is relevant to the
$P=1atm$ phenomenon.  An extensive   databank for these  properties is
not yet available,  but some relevant pieces  are. From the  available
experimental data,  $\eta$ and $\tau_\alpha$ are  seen to  change typically by
$8$ orders of magnitude along the isochores whereas a mere increase of
$1$ order of magnitude is observed along isotherms:  see Figs.  2 to 5
of Ref.\cite{FLDKAT98} and Fig.  4  of Ref. \cite{LKOHC95}. A detailed
study of this   phenomenon  is given  in   Ref. \cite{FLDKAT98}.   One
concludes that    density  seems to  play a    minor role compared  to
temperature, which  indicates   that   the viscous slowing  down    of
supercooled  liquids, at  least at  low  pressure, does not  primarily
result from a congestion  mechanism  such as described by  hard-sphere
models and simple free-volume theories.

\section{Simple atomistic models versus real fragile glassformers}

Liquid   models involving  one   or   two components  of   spherically
interacting particles, such as the binary Lennard-Jones system, are at
the basis of many computer simulation  studies and various approximate
theoretical  treatments.   It    is  implicitly   assumed  that    the
characteristics  of  their  viscous  slowing down  as   temperature is
lowered   reproduce  the main features  of   the phenomenology of real
glassforming liquids. We  focus here on the  ability of such models to
describe the marked superArrhenius activated  behavior of $\alpha$ relaxation
and viscosity that, as stressed above, is  a distinct property of real
fragile supercooled liquids.  

The most straightforward way to  examine  the relevance of the  simple
models is  to plot the effective  activation  free energy $E(T)$  as a
function of the scaled temperature $T/T^*$, or its inverse. To minimize
bias, we use the values of $E(T)$ {\it and} $T^*$ given by Sastry {\it et al}\cite{SDS98}
in their  study   of the correlation between   the  properties of  the
inherent   structures  (potential  energy   landscape)  and   the slow
relaxation. (Note that on the basis of their plot of $E(T)$ versus
$T/T^*$, with $T^*=1$ in Lennard-Jones reduced units, Sastry {\it et
al}\cite{SDS98} pointed out that the binary Lennard-Jones model behaves as a
fragile liquid.)   The  plots are displayed in   Fig.  2a, where  $E(T)$ is
measured in units of $k_BT^*$ and in Fig.   2b where the deviation from
the high-$T$ Arrhenius value,  ($E(T)-E_\infty  )/k_BT^*$, is shown.    The
result  is  striking: the binary   Lennard- Jones system  appears as a
strange hybrid.  On one hand,  it has very weak intermolecular bonding
that leads to  a small value of  the high-$T$ Arrhenius  activation free
energy $E_\infty /k_BT^*\simeq 2.45$ characterizing ``normal'' liquid behavior,
in  contrast to weakly-bonded ortho-terphenyl  for which  the ratio is
$9$, to  intermediate H-bonded glycerol for  which the ratio  is $18$,
and to the strong network-forming $GeO_2$ for which the ratio is $27$;
on the other  hand, its departure  from Arrhenius behavior below $T^*$,
when properly scaled to be compared to real glassforming liquids as in
Fig. 2b, is comparable to that of  the {\it strongest} glassformers, such as
$GeO_2$, and is  much less  than that of  fragile liquids  like ortho-
terphenyl\footnote{The  relevant  simulations  were  performed at {\it
constant  volume}, and  the constant-pressure behavior   of the binary
Lennard-Jones model has not been  checked yet.  It should be stressed,
however,  that  a strong influence  of density,   resulting in a major
difference between constant-pressure and constant-volume properties of
the binary Lennard-Jones system, would be at odds with the behavior of
real  supercooled liquids:  see section  2.}.   We conclude that  the
binary Lennard-Jones model  appears  to be  a very weakly-bonded,  but
nonfragile (in the sense of showing only weak superArrhenius behavior)
system. This conclusion, that contradicts the generally accepted view
about the fragility of Lennard-Jones systems (including  the recent suggestion, based on  a rather indirect argument, by
Angell et al \cite{ARV99} that the binary Lennard-Jones
mixture is a moderately fragile liquid), stresses the importance of a
proper scaling of the available data. In the case of Lennard-Jones models,
the conventional use of $T_g$ as a scaling temperature is questionable
because the determination of $T_g$ requires an uncontrolled extrapolation
procedure, with molecular-dynamics simulation data stopping some 10 or 11
orders of magnitude short of the typical $\alpha$-relaxation time at the
calorimetric $T_g$.

Other systems like  one-component liquid models with atoms interacting
through the  Lennard-Jones  and the Dzugutov\cite{D92} potentials  are
also  very   nonfragile.     (models   of  molecular   liquids    like
ortho-terphenyl are not considered  here.)  Anticipating  the analysis
in the next  section, we quantify the  ``fragility'' of a system  by a
parameter $B$ that measures  its  degree of departure from   Arrhenius
behavior (in a plot of ($E(T)-E_\infty )/k_BT^*$ versus $T/T^*$): the more
fragile  a system, the  more superArrhenius the  $T$-dependence of its
viscosity and $\alpha$-relaxation time,  the larger $B$. For instance, the
value of $B$   is found to be  more  than $400$ for   the very fragile
ortho-terphenyl whereas it is equal to $28$ for the strong glassformer
$GeO_2$,    and    to   $90$    for    the   intermediate       system
glycerol\cite{KTZK96}; in contrast we have found $B\simeq 1.0$, $1.3$, and
$5.4$  for  the  one-component  Lennard-Jones,  the Dzugutov,  and the
binary Lennard-Jones systems, respectively\footnote{In the language of
the frustration-limited domain  approach   (see  section 4),   $B$  is
inversely  related  to   the   strength  of   the inherent  structural
frustration in the  system.  The small values   of $B$ for  the simple
atomistic models composed of one or two species of spherical particles
indicate that they are strongly frustrated.  This can also be inferred
from the work of  Nelson and coworkers\cite{SN84} on metallic glasses,
where  it was found  that the density of  defects  associated with the
frustration of icosahedral order is high,  so that correlation lengths
never     grow      significantly      beyond      the     interatomic
distance.}\cite{FSKA99,TAFSK99}, values  characteristic      of   very
nonfragile   liquids.  The values  obtained    for  $B$ are   somewhat
dependent  upon the fitting procedure,   since the range of  available
relaxation times is much  more   limited for these computer    studied
systems than for real glassforming liquids; however, the fact that the
obtained values  of  $B$ are  typical of nonfragile  liquids, with
$B \leq 20$, seems robust.

Supportive  of  the above  conclusion are  also the recent  studies of
one-component Lennard-Jones systems   in which the slow  relaxation is
described  kinetically by means   of  an appropriate master  equation,
acting in a  potential  energy landscape\cite{APRV98,MDW99}.  Although
finite-system-size effects, incomplete information on the connectivity
between potential  energy minima, and the  use of other approximations
in handling the master equation approach, can still  cast doubt on the
validity and the generality of the results, one notices the remarkable
finding  of  these studies,  that the   temperature dependence  of the
viscosity  and relaxation     time   of  Lennard-Jones    systems   is
Arrhenius-like  over the whole range of  up to 13 decades in viscosity
and time, a property suggestive of a nonfragile liquid! See Fig. 3a of
Ref.\cite{APRV98}  and Fig. 10   of Ref.\cite{MDW99}.   The discussion
above    does not, of   course, diminish   interest   in the study  of
relaxations in simple    atomistic  models per se;  but    it strongly
questions the  relevance  of such studies   in  shedding light  on the
dominant mechanisms that lead to  the glass transition of real fragile
supercooled liquids.

\section{Frustration-limited domain theory}

The description   of   the  viscous slowing down     of  weakly-bonded
glassforming   liquids  as a   superArrhenius activated   process with
sizeable activation  free energies  is  indicative of the  presence of
collective  or cooperative phenomena;  in addition,  the heterogeneous
nature    of the  $\alpha$   relaxation  suggests   that this  collective  or
cooperative character shows up in  supercooled liquids in the form  of
spatial  domains,   supermolecular    regions over   which   dynamical
properties  are correlated.  Numerous models have  been   based on the
existence of domains, clusters, fluctuations, cooperative regions, and
2- liquid phases\cite{S99}.  While providing a reasonable  way  of fitting or
rationalizing pieces of  the phenomenology, these models usually leave
unanswered the questions  of why heterogeneities  or domains appear in
supercooled liquids and how  they account for  the activated nature of
the  $\alpha$   relaxation?  The  theory   of frustration-limited domains\cite{KKNT95}
addresses these  basic questions.  It is a  nonmolecular, ``mesoscopic''
approach, in which the driving  force for domain formation and  growth
comes  from an ordering transition that  is  aborted by the ubiquitous
presence of structural frustration. We summarize this approach in what
follows.

\subsection{Physical picture}
The basic   physical  ingredient   is   the  concept    of  structural
frustration.  Molecules in a liquid,  even in the "normal" range above
melting, tend to arrange themselves into a locally preferred structure
corresponding to the minimization of an appropriate local free energy;
but the  spatial extension  of this local  arrangement  is thwarted by
ubiquitous  structural frustration that  prevents a periodic tiling of
space by the locally preferred structure and is the source of a strain
free energy  that  grows super-  extensively with   system  size. As a
result of  the competition between  the short-range tendency to order,
i.   e., to  extend the locally   preferred structure, and the  strain
generated  by  frustration,  the liquid, below   some  temperature $T^*$,
breaks up into  domains whose structure  is the locally preferred, but
strained, one associated  with the high-$T$ liquid,  and whose  size and
further  growth are limited  by  frustration.  $T^*$ is  interpreted as a
crossover between  normal,  molecular   liquid behavior   ($T>T^*$)   and
collective, domain-influenced behavior  ($T<T^*$).  As seen below, in the
theory $T^*$   does indeed  mark   the   passage from Arrhenius-like   to
superArrhenius $T$-dependence  of  the  $\alpha$-relaxation  time and viscosity
that we have stressed in the preceding sections.

\subsection{Theoretical model}
The  above  picture  can be  incorporated  in a
coarse-grained  model based   on  an  effective Hamiltonian\cite{KKNT95}.   The
locally preferred structure of the  liquid is characterized by a local
order variable ${\bf O(r)} $ that can take a discrete set of orientations (more
than $2$), as in a clock model. The Hamiltonian  consists of a reference
term describing   the short-range tendency   to order and  a weak, but
long-range  competing        interaction        that   generates   the
frustration-induced strain:
\begin{equation}
H=-\sum_{{\bf r},{\bf r'}}J({\bf r}-{\bf r'}){\bf O(r)}{\bf .} {\bf O(r')}+\frac{Qa_0}{2}
\sum_{{\bf r},{\bf r'}}\frac{{\bf O(r)}{\bf .}{\bf O(r')}}{|{\bf r}-{\bf r'}|},
\end{equation}
where $J({\bf r})$ is a short-range ``ferromagnetic'' interaction with
a  typical energy  scale  $J$,  $a_0$  is  the characteristic   length
associated with the   local structure  (of the    order  of the   mean
intermolecular distance in a liquid),  and $0<Q<<J$. In the absence of
frustration, the  reference  system  undergoes  a  continuous ordering
transition at a temperature $T^*\sim J$.

\subsection{Avoided  critical  behavior}
Due to the long-range  nature of the  competing interaction that gives
rise  to a superextensive cost  in  free energy proportional to $L^5$,
long-range order characteristic  of the reference, unfrustrated system
(here,  ferromagnetic order) is  completely prohibited by frustration,
no matter  how small, but nonzero,  this latter may be\cite{KKNT95}. A
still stronger  statement can be made about   the system's behavior in
the  temperature-frustration  ($Q/J$)  phase   diagram.   An  ordering
transition,  leading to      a    phase with    modulated  order    (a
``defect-ordered phase''),   can   take place at   a   low temperature
$T_{DO}(Q)<T^*$,  but  the critical  point  at  ($T^*, Q=0$)  is  {\it
isolated}, i. e., no other critical points are present in its vicinity
in  the $T-Q$ phase diagram.  This  property has been shown rigorously
for  the  spherical version of   the model\cite{CEKNT96}  and  for the
$O(n)$  version\cite{NRKC99}  with    $n$   strictly     larger   than
two\footnote{For Ising spins  ($n=1$),   the transition line  to   the
defect-ordered  phases goes  continuously, albeit nonanalytically,  to
the critical point at $T^*$ when  $Q\to 0$.  However, the transition in
the   presence of  frustration   is  not  continuous:   it   is driven
{\it first-order}   by the fluctuations\cite{VT98}.},   for  which $lim_{Q\to
0}T_{DO}(Q)$ is significantly  less than $T^*$.  For  weak
frustration, $Q<<J$,  the  critical point at  $T^*$  is  {\it narrowly
avoided}, which results in the  presence of two supermolecular lengths
below $T^*$:  the correlation length $\xi_0$  of  the reference system,
$\xi_0\sim (T^*-T)^{-\nu}$   with   $\nu\simeq     2/3$  for   all   standard
$3$-dimensional critical systems in  the absence of quenched disorder,
and  a     new    length, $L^*\sim (Q/J)^{-1/2}\xi_0^{-1}\sim (Q/J)^{-1/2}(T^*-
T)^{+\nu}$, that is the characteristic scale  over which the physics is
essentially  that of the unfrustrated  system (which is ordered, since
$T<T^*$)  and that  can  be    identified  with the typical     domain
size\cite{KKNT95}.

\subsection{Static  and   dynamic  phenomenological  scaling
analysis} 
At temperatures sufficiently  below $T^*$, so that $a_0<< \xi_0<<L^*$,
one  can     develop scaling arguments    to  describe  the collective
contribution   to both static and   dynamic properties.  The liquid is
considered    as     a   collection  of     (slightly)   polydisperse,
frustration-limited domains   whose   mesoscopic order  parameters are
randomly  oriented.   The static  scaling   analysis  starts  with the
formulation  of the  free-energy density of  a  domain of size $L$  in a
random environment\cite{KKNT95},
\begin{equation}
\frac{F(L)}{L^3}=\frac{\sigma}{L}-\phi+sL^2,
\end{equation}
where, as in a standard nucleation picture, the surface tension ($\sigma$)
scales as $\xi_0^{-2}$ and the ordering bulk free-energy density ($\phi$)
as $\xi_0^{-3}$; the  additional term describes the frustration-induced
strain, whose coefficient $s$ is {\it a priori} unknown. Assuming that
the {\it    inter}-domain interactions are   sufficiently  weak  to be
treated by a mean-field approach,  one can derive  the scaling form of
the strain coefficient  $s$ by identifying  the length that  minimizes
the free-energy density with   the typical domain size  $L^*$,  namely
$s\sim  \sigma/(L^*)^3\sim (Q/J)^{3/2}\xi_0$;  this  leads  to the  equilibrium
distribution of domain sizes, $\rho (L,T)$, in a scaled form,
\begin{equation}
\rho(L,T)\propto
\exp\left[-\gamma(T)\left(\kappa\left(\frac{L}{L^*}\right)^2-\frac{3}{2}\left(\frac{L}{L^*}\right)^3+\frac{1}{2}
\left(\frac{L}{L^*}\right)^5\right)\right],
\end{equation}
where the entire temperature dependence is  contained in the parameter
$\gamma(T)\sim (Q/J)(T^*/T)(1-T/T^*)^{8/3}$ and $\kappa$  is a number of order  1.

The presence  below $T^*$ of internally  ordered  domains leads to slow,
activated dynamics of the  (discrete) order variable.  By generalizing
simple  dynamic scaling   arguments     about slow relaxation   in   a
finite-size  system below   its  ordering transition  via  defect-wall
creation,   one    can evaluate   the    collective (or   cooperative)
contribution to  the activation free  energy, $\Delta E(L,T)$, for a domain of
size $L$; this latter takes the following scaled form\cite{KKNT95},
\begin{equation}
\frac{\Delta E(L,T)}{k_BT}=b\gamma(T)\left(\left(\frac{L}{L^*}\right)^2-m\left(\frac{L}{L^*}\right)^5\right),
\end{equation}
where $b$ and  $m$ are  numbers of  order  1, with $bm<1/2$.  The full
effective  activation  free energy  is then   obtained by   adding the
molecular,  Arrhenius-like contribution, $E_\infty   /k_BT$, to the  above
expression.    It  is  important  to  stress  that   the  domains  are
equilibrium  entities  (in  a supercooled liquid  they   are of course
metastable  relative  to the crystal),  and   that the slow, activated
dynamics  does {\it    not} correspond to   out-of-equilibrium  domain
coarsening.

\subsection{Main predictions}
 The  central point  of the  theory  concerns the
prediction of a crossover  from  molecular, normal liquid  behavior to
collective,  domain-dominated  behavior  around  $T^*$,  a crossover that
explains the change from Arrhenius-like to superArrhenius $T$-dependence
of   the    viscosity  and   $\alpha$-relaxation    time. The  superArrhenius
contribution to the activation free energy is described by a universal
power law, i. e.,
\begin{equation}
\begin{array}{cllc}
E(T)&=& E_\infty & {\rm for }\,\, T>T^*,\nonumber\\
&=&E_\infty+B(k_BT^*)\left(1-\frac{T}{T^*}\right)^{8/3}& {\rm for }\,\, T<T^*,
\end{array}
\end{equation}
a prediction  that fits     well  the  extant experimental  data    on
glassforming liquids.   A central  feature of the  theory is   that it
predicts  the nontrivial, and   somewhat  unusual, universal  exponent
$8/3$,     which  has     been     reasonably    well  verified     by
experiment\cite{KTZK96}.  For an illustration,  see fig.  2b as well as
Refs.\cite{KTZK96} and \cite{KKNT95}.  The parameter $B$ in Eq. 6 is a
measure   of  the   departure  from Arrhenius    behavior,    hence of
``fragility'', and it is the one we have used in the preceding section
about  the  relevance of  simple atomistic systems.  From  the scaling
analysis,  we obtain that $B\sim (Q/J)^{-1}$,   so that the more fragile
(superArrhenius) a liquid, the less frustrated.

The dynamics are  also naturally described  as heterogeneous below the
temperature $T^*$  at which the  liquid  breaks up into supermolecular
domains.  For instance, the nonexponential character of the dielectric
relaxation  can  be   simply  modeled   by  assuming  that   molecular
reorientations are completed within a single domain, in which they are
dynamically  coupled to the order  variable, and by  assuming that the
relaxation within each  domain is exponential  in time.  This leads to
the  following expression for  the (normalized)  dielectric relaxation
function\cite{KKNT95}:
\begin{equation}
f_\alpha(t)=\int_0^\infty dL L^2\rho(L,T)\exp\left[-\frac{t}{\tau_{\alpha,\infty}
}\exp\left(-\frac{E_\infty +\Delta E(L,T)}{k_BT}\right)\right],
\end{equation}
for which $\rho(L,T$)   and $\Delta  E(L,T)$ are  given  by the  theory.  As
illustrated in Fig.  3, by comparing to  the experimental data  on the
imaginary part of the frequency-dependent dielectric susceptibility of
the fragile  glassformer salol\cite{DWNWC90},  the above  equation (or
rather the  Fourier transform  of  $df_\alpha (t)/dt$), together  with the
scaling expressions for $\rho (L,T)$ and $\Delta E(L,T)$ given in Eqs. 4 and
5, allow one to reproduce  quite well the   main features of the  $\alpha$
relaxation     with   a  small       set of    temperature-independent
parameters. More details will be given in a forthcoming publication.

The  relevance  of   the  frustration-limited  domain theory   to  the
description of other   aspects of  the  phenomenology  of  supercooled
liquids,  such as  the   decoupling of  translational and   rotational
relaxation, the avoidance   of the Kauzmann entropic  catastrophe, and
the possible   observation   of   ``defect-ordered  phases''   at  low
temperature, is discussed elsewhere\cite{TK95}.

\subsection{ Open  questions}
A fundamental  theoretical uncertainty yet  to be resolved  is whether
the assumptions  that underlie  the phenomenological scaling  analysis
leading  to  activated, heterogeneous  dynamics evolve rigorously from
the  Hamiltonian in Eq. 2 (completed   by some short-time dynamics for
the spin variables).  A more physical problem is  that  of mapping the
local order and  the frustration in  a given liquid, however they  are
defined, onto the  local order variable, $  {\bf O(r)}$,  and onto the
frustration parameter $Q$ (or equivalently onto $B$).

\section{Conclusion}

In  this article, we stress  that the salient  feature to be explained
about     the    glass  transition     is  the  temperature-controlled
superArrhenius activated   nature   of the viscous   slowing   down of
supercooled liquids,  more strikingly seen  in weakly-bonded,  fragile
systems. In  the light   of this observation,  we  have   compared the
behavior   of simple  models   involving  one or   two  components  of
spherically interacting atoms with  that of real glassforming  liquids
and shown that, although  being very weakly interacting systems, these
simple models are at the same time nonfragile, in the sense that their
viscosity and  $\alpha$-relaxation time show  little departure from Arrhenius
temperature dependence.  Finally, we have  summarized how our approach
to  supercooled  liquids, based  on  frustration-limited domains,  can
explain the main aspects of the phenomenology, including the crossover
from Arrhenius   to   superArrhenius behavior and   the  heterogeneous
character of the $\alpha$ relaxation.

\section*{Acknowledgements}

This work has been supported by  the CNRS, the NSF,  and NATO. We wish
to thank  Dr. Christiane Alba-Simionesco and  Prof. Steven A. Kivelson
for many stimulating discussions.

\newpage

\begin{figure}\label{fig:1}
\begin{center}


\caption{ SuperArrhenius T-dependence of the viscosity $\eta$ and
$\alpha$-relaxation time $\tau_\alpha$  in several representative glass-forming
liquids.  a) Logarithm  (base 10) of $\eta$  and $\tau_\alpha$  versus reduced inverse
temperature $T_g/T$  for  $GeO_2$,  a system  forming  a  network  of strong
intermolecular  bonds,  ortho-terphenyl,   a  weakly-bonded  molecular
liquid, and  glycerol, an ``intermediate'' hydrogen-bonded liquid. For
the former,  the variation is  almost linear (Arrhenius-like), whereas
the two others  are  characterized, below  some temperature  $T^*$, by  a
strong  departure  from  linear dependence  (superArrhenius behavior).
(Data  taken  from   references cited in   Ref.\cite{KTZK96}.)  b)  Effective
activation free  energy  $E(T)$, obtained from  data shown  in a), as  a
function of inverse  temperature.  Both $E(T)$ and $T$ are scaled  by the
crossover temperature $T^*$ shown in a). ($T^*=350K, 322K, 1150K$ for ortho-terphenyl, glycerol, and $GeO_2$, respectively.)}
\end{center}
\end{figure}

\begin{figure}\label{fig:2}
\begin{center}

\hfill

\caption{
Relative  ``fragility''  (superArrhenius character)   of  the   binary
Lennard-Jones model compared to   a strongly-bonded, a  weakly-bonded,
and  an intermediate supercooled liquid.  a) Effective activation free
energy $E(T)$ for  $GeO_2$, glycerol,  orthoterphenyl, and the  binary
Lennard-Jones model,  as a function  of temperature.   Both $E(T)$ and
$T$ are scaled by  the crossover temperature  $T^*$. Note that the data
for $E(T)$  as well as the value  of $T^*$ for the binary Lennard-Jones
model (at constant volume) are directly taken from Ref.\cite{SDS98}. b) Same as
in a),  but  with the  high-$T$ Arrhenius  contribution subtracted for
each system. The dotted lines represent the 8/3 power-law prediction
of the frustration-limited domain theory, with the values of the
(adjustable) parameter B taken from Refs.\cite{KTZK96} and \cite{TAFSK99} and given in the
text.}
\end{center}
\end{figure}

\newpage
\begin{figure}\label{fig:3}
\begin{center}
\hbox to\hsize{\epsfxsize=0.6\hsize\hfil\epsfbox{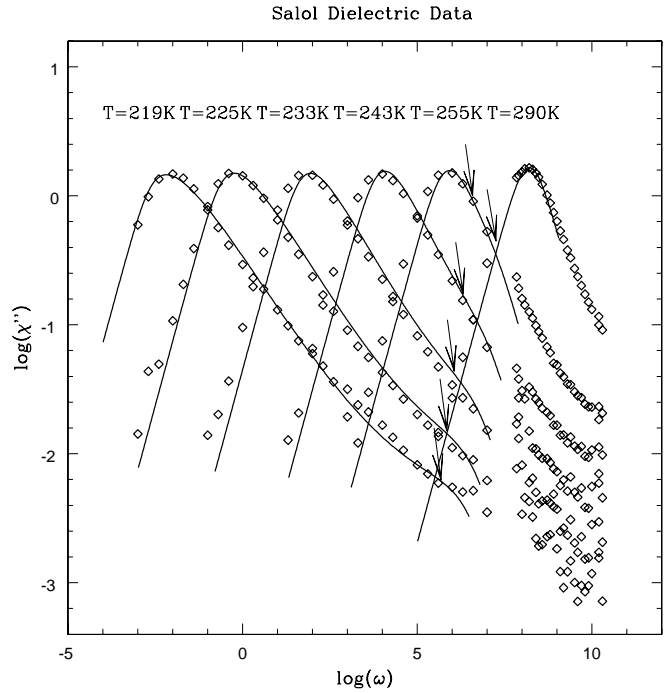}}
\hfill

\caption{ 
Comparison  between the frequency-dependent dielectric  susceptibility
of  the  fragile  glassformer   salol  and the   predictions   of  the
frustration-limited domain    theory,    Eqs.    4,    5,  and      7;
$\log_{10}(\chi''(\omega  ))$  is plotted versus $\log_{10}(\omega)$  at several
temperatures.  The  temperature-  and frequency-independent parameters
$\kappa$,  $b$, and  $m$ are  obtained by  fitting  the  data: $\kappa =0.84$,
$b=1.46$,   $m=0.20$.   The    other   parameters    are  taken   from
Ref. \cite{KTZK96}, and the experimental data are taken from Ref. \cite{DWNWC90}.
Due  to its  nonmolecular, mesoscopic   character, the  theory has  an
intrinsic   upper-frequency    cut-off,   $\omega_{cut}\sim (\omega_\infty/10)\exp(-
E_\infty/k_BT)$, marked by the arrows.}
\end{center}
\end{figure}
\end{document}